\documentclass[useAMS,usenatbib]{mn2e}

\usepackage{graphicx}
\usepackage{amsmath}
\usepackage{amssymb}
\usepackage{color}
\usepackage{subfig}
\usepackage[breaklinks,colorlinks,citecolor=blue,linkcolor=red]{hyperref} 
\usepackage[all]{hypcap}

\newcommand\msun{\, \rm M_\odot}

\newcommand\kms{\, \rm km\,s^{-1}}

\newcommand\aout{{a_{\rm out}}}
\newcommand\ain{{a_{\rm in}}}
\newcommand\vdisp{{\sigma_{\rm disp}}}

\title[A dynamical origin for planets in triple stars]{A dynamical origin for planets in triple star systems}
\author[G. Fragione, A. Loeb, I. Ginsburg]{Giacomo Fragione$^{1}$\thanks{E-mail: giacomo.fragione@mail.huji.ac.il}, Abraham Loeb$^{2}$, Idan Ginsburg$^{2}$\\
$^{1}$Racah Institute for Physics, The Hebrew University, Jerusalem 91904, Israel\\
$^{2}$Astronomy Department, Harvard University, 60 Garden St., Cambridge, MA 02138, USA}

\begin{document}

\maketitle

\begin{abstract}
Recent radial velocity and transit data discovered $\sim 100$ planets in binary or triple stellar systems out of the entire population of a few thousand known planets. Stellar companions are expected to strongly influence both the formation and the dynamical evolution of planets in multiple star systems. Here, we explore the possibility that planets in triples are formed as a consequence of the dynamical interactions of binaries in star clusters. Our simulations show that the probability of forming a planet-hosting triple as a consequence of a single binary-binary scattering is in the range $0.5-3\%$, when one of the binaries hosts a planet. Along with other formation scenarios, binary-binary encounters are a viable way of creating planet-hosting triple systems. The recently launched TESS satellite is expected to find a larger sample of planets in triple systems, and to shed light on their origin.
\end{abstract}


\section{Introduction}

The discovery of new exoplanets has accelerated over the past decade through observations with the transit and Doppler technique. Approximatively $3700$ exoplanets have been confirmed\footnote{https://exoplanet.eu}, most of which discovered by the \textit{Kepler} satellite\footnote{https://\rm www.nasa.gov/mission\_pages/kepler/main/index.html} \citep{burk2014,cou16,math17}. The recently launched transit mission TESS\footnote{https://tess.gsfc.nasa.gov/} will monitor thousands Sun-like stars for nearly across the entire sky and is expected to observe a population of exoplanets that is on order of magnitude larger than \textit{Kepler}'s population \citep{ricker}.

More than one hundred of the known exoplanets have been found in binary or higher-order systems. Only two dozens of triple stars and a couple of quadruple systems have been observed to host a planet\footnote{http://www.univie.ac.at/adg/schwarz/multiple.html}. Several observational campaigns have shown that stars tend to be born in pairs and higher-order hierarchies. More than $\sim 40$\%-$50\%$ of stars are thought to have at least one stellar companion, while $\sim 20\%$ are observed in triple or higher-order systems \citep{duq91,rag10,tok14a,tok14b}. As a consequence, a natural question arises as to whether the paucity of planets observed in multiple systems is due to observational biases or induced by the gravitational effect of the multiple stellar companions on the planet's dynamics.

\begin{table*}
\caption{Multiplicity of Solar-like stars and exoplanet host stars.}
\centering
\begin{tabular}{ccccc}
\hline
Single & Binary & Triple or higher & Multiple (N$>1$) & Ref.\\
\hline\hline
\multicolumn{5}{c}{Solar-like stars}\\
\hline
$54$ \% & $34$ \% & $12$ \% & $46$ \% & \citet{rag10} \\
$56$ \% & $38$ \% & $6$  \% & $44$ \% & \citet{duq91} \\
$41$ \% & $46$ \% & $13$ \% & $59$ \% & \citet{tok14a,tok14b} \\
\hline\hline
\multicolumn{5}{c}{Exoplanet host stars}\\
\hline
$77.1$ \% & $19.8$ \% & $3.1$ \% & $22.9$ \% & \citet{rag06} \\
$82.8$ \% & $14.8$ \% & $2.4$ \% & $17.2$ \% & \citet{mug09} \\
$88.0$ \% & $9.9$  \% & $2.1$ \% & $12.0$ \% & \citet{roe12} \\
\hline\hline
\end{tabular}
\label{tab:obs}
\end{table*}

\begin{table*}
\caption{Models: label, mass of stars in first binary ($m_1=m_2$), mass of stars in second binary ($m_3=m_4$), semi-major axis of first binary ($a_1$), semi-major axis of second binary ($a_2$), eccentricity of first binary ($e_1$), eccentricity of second binary ($e_2$), velocity dispersion ($\vdisp$), planet's semi-major axis ($a_P$).}
\centering
\begin{tabular}{lcccccccccc}
\hline
Name & $m_1=m_2$ ($\msun$) & $m_3=m_4$ ($\msun$) & $a_1$ (AU) & $a_2$ (AU) & $e_1$ & $e_2$ & $\vdisp$ ($\kms$) & $a_P$ (AU) \\
\hline
Model 1   & $1$;$3$;$5$   & $1$ 	  	  & $10$;$20$;$50$;$100$  & $10$;$20$;$50$;$100$ & $0$ 		   			   & $0$ 		 & $3$       & $1$ \\
Model 1b  & $1$       	  & $1$;$3$;$5$   & $10$;$20$;$50$;$100$  & $10$;$20$;$50$;$100$ & $0$ 		   			   & $0$ 		 & $3$       & $1$ \\
Model 2   & $1$;$3$;$5$   & $1$       	  & $10$ 	    		  & $10$ 		 		 & $0$;$0.2$;$0.4$;$0.6$   & $0$ 		 & $3$       & $1$ \\
Model 3   & $1$;$3$;$5$   & $1$       	  & $10$ 	    		  & $10$ 		 		 & $0$ 		   			   & $0$         & $3$       & $0.1$;$0.5$;$1$ \\
Model 4   & $1$	          & $1$       	  & $10$;$20$;$50$;$100$  & $10$;$20$;$50$;$100$ & $0$ 		   			   & $0$ 		 & $0.1$;$0.3$;$0.5$;$1$;$2$;$3$ & $1$ \\
\hline
\end{tabular}
\label{tab:models}
\end{table*}

Planet searches usually target single stars and very wide binaries since close companions make the data analysis more complicated. Follow-up observations have also revealed planets in close binary systems \citep{desid07,thh15}. In some cases, one of the stars appeared to be part of a binary system making the host a hierarchical triple system. To reveal companions to the host star, high-angular resolution observations are needed to separate contaminating nearby stars that lie within the aperture. This makes it challenging to conduct follow-up observations for both confirmed and candidate exoplanetary systems \citep{fur17,hirs17,zie18}. While a simple calculation of the relative fraction of planets in triples leads to $\sim 0.7\%$ of the population, some follow-up observations on a sub-sample of known planets have suggested this fraction can be as large as $\sim 2-3\%$ \citep{rag06,mug09,roe12}. Table \ref{tab:obs} reports a collection of observations of stellar multiplicity and planet-host stellar multiplicity. Note that the relative fraction of binaries is defined as $N_b/(N_s+N_b+N_{th})$, where $N_s$, $N_b$, and $N_{th}$ are the number of single, binary and triple (or higher order) systems, respectively. Similarly, the relative fraction of triples (or higher order) systems is defined as $N_{th}/(N_s+N_b+N_{th})$.

The nearest star system to the Sun is a triple made of the binary $\alpha$ Centauri with an outer companion, Proxima Centauri \citep{kerv17}, hosting a terrestrial-mass planet in Proxima's habitable zone \citep{anglada}. Stellar companions could have a profound influence on the planet formation process, truncating disks and enhancing accretion and photoevaporation, as well as the dynamical instability, e.g. due to Kozai-Lidov resonances \citep{naoz13,dut16}. 

The gravitational influence of stellar companions is expected to strongly affect planetary system formation \citep{wang14}. Nevertheless, the difficulty in identifying planets in binary and multiple systems has left the significance of this effect uncertain. While planet formation and stability in binary stars has been investigated in several papers \citep{naoz13,kraus16}, little attention has been dedicated to planets in triple stars due to the absence of a large enough sample. Four different channels can make planets in triple systems: binary-binary scatterings \citep{port05}, primordial triple formation \citep{domin15}, capture of planets by a triple \citep{peret12}, and capture of stars with planets by binaries \citep{moec11}. In this paper, we consider the dynamical origin of planets in triple systems as a consequence of binary-binary scatterings in star clusters, as first proposed by \citet{port05}. Binary-binary encounters may dominate the cluster dynamics even for a relatively moderate fraction of stars in binaries \citep{lei13}, which might be common in the majority of clusters \citep*{frag17}. For the first time, we directly include a planet in the scattering experiments. Our planet is originally bound to a star in one of the two binaries. We consider different masses, orbital semi-major axis and eccentricities of the stars in the binary, and also different values for the planetary semi-major axis. Finally, we study how the diverse local environments affect the dynamical formation of planets in triple systems.

The paper is organised as follows. In Section 2, we describe the methods and initial conditions we used in our scattering experiments. In Section 3, we present the results of our $N$-body calculations. Finally, we discuss the implications of our findings in Section 4.

\section{Method}

Our scattering experiments were performed primarily using \textsc{fewbody}, a numerical toolkit for simulating small-$N$ gravitational dynamics \citep{fregeau04}. In total, we ran 25k scattering experiments for each set of initial conditions for a total of $\sim 2.6$ million simulations.

The initial conditions of our runs are summarized in Table \ref{tab:models}. We consider equal mass binaries (i.e. $m_1=m_2$ and $m_3=m_4$) made up of $1$-$3$-$5 \msun$ stars with initial semi-major axis ($a_1=a_2$) between $10$ AU and $100$ AU, and initial eccentricities $e_1=0$-$0.6$ and $e_2=0$. We fix the mass of the planet to Jupiter mass and vary its initial semi-major axis in the range $0.1$-$1$ AU. Finally, we fix the relative velocity of stars to the velocity dispersion of the host cluster, which we vary in the range $\vdisp=0.1$-$3\kms$. We sample the impact parameter from a distribution
\begin{equation}
f(b)=\frac{b}{2b_{\rm max}^2}\ ,
\end{equation}
where $b_{\rm max}$ is the maximum impact parameter defined by
\begin{equation}
b_{\rm max}=p_{\rm max}\sqrt{1+\frac{2G M_T}{p_{\rm max}\sigma_{\rm disp}^2}}\ ,
\end{equation}
where $M_T=m_1+m_2+m_3+m_4+m_P$ is the total mass of the system, $\vdisp$ is the velocity dispersion, and $p_{\rm max}$ is the maximum pericentre distance of the binary-binary encounter. We set $p_{\rm max}=5(a_1+a_2)$ \citep{heg96}.

Six angles describe the relative phases and orientations of the system. Given the plane of motion of the two binaries centres of mass, the relative inclinations of the orbital planes of the two binaries constitute the first set of angles. The initial relative phases of the stars in each binary add two more angles. Finally, the inclination of the planet orbit with respect to the host binary orbital plane and the initial phase of the planet along its orbit yield the last two angles. For all our scattering experiments, these phase and orientation angles are chosen randomly, so that the outcomes represent averages over these quantities. 

We choose the initial separation of the two systems to be the distance at which the tidal perturbation on each system has a fractional amplitude $\delta=F_{\rm tid}/F_{\rm rel}=10^{-5}$, where $F_{\rm tid}$ and $F_{\rm rel}$ are the initial tidal force and the relative force between each component of the system, respectively \citep{fregeau04,antogn16}. Ignoring the planet, there are four outcomes of a binary-binary encounter: (i) two binaries (2+2); (ii) a triple and a single star (3+1); (iii) a binary and two single stars (2+1+1); (iv) complete ionizations (1+1+1+1). The latter outcome is possible, but extremely rare \citep{lei16}. When the planet is taken into consideration, it can be bound to a star in one of the above hierarchies or can be ejected being unbound to any of them. \textsc{fewbody} classifies the N-body system into a set of independently bound hierarchies and considers a run completed when their relative energy is positive and the tidal perturbation on each outcome system is $\le \delta$. Moreover, \textsc{fewbody} checks the stability of the outcomes through the \citet{mar01} stability criterion for hierarchical triple systems,
\begin{equation}
\frac{\aout}{\ain}>\frac{2.8}{1-e_{\rm out}}\left[\left(1+\frac{m_{\rm out}}{m_{\rm in}}\right)\frac{1+e_{\rm out}}{\sqrt{1-e_{\rm out}}}\right]^{2/5}\left(1-\frac{0.3 i}{180^\circ}\right)\ ,
\end{equation}
where $m_{\rm in}=m_{\rm in,1}+m_{\rm in,2}$ is the total mass of the inner binary, $m_{\rm out}$ is the mass of the outer companion and $e_{\rm out}$ its orbital eccentricity, and $i$ is the relative inclination between the inner and outer orbit. In the case of a hierarchical quadruple system, as in the case of a triple containing a planet, we first apply the triple stability criterion to the inner triple, then to the outer triple, approximating the innermost binary as a single object \citep{fregeau04}.

\section{Results}

\begin{figure} 
\centering
\includegraphics[scale=0.58]{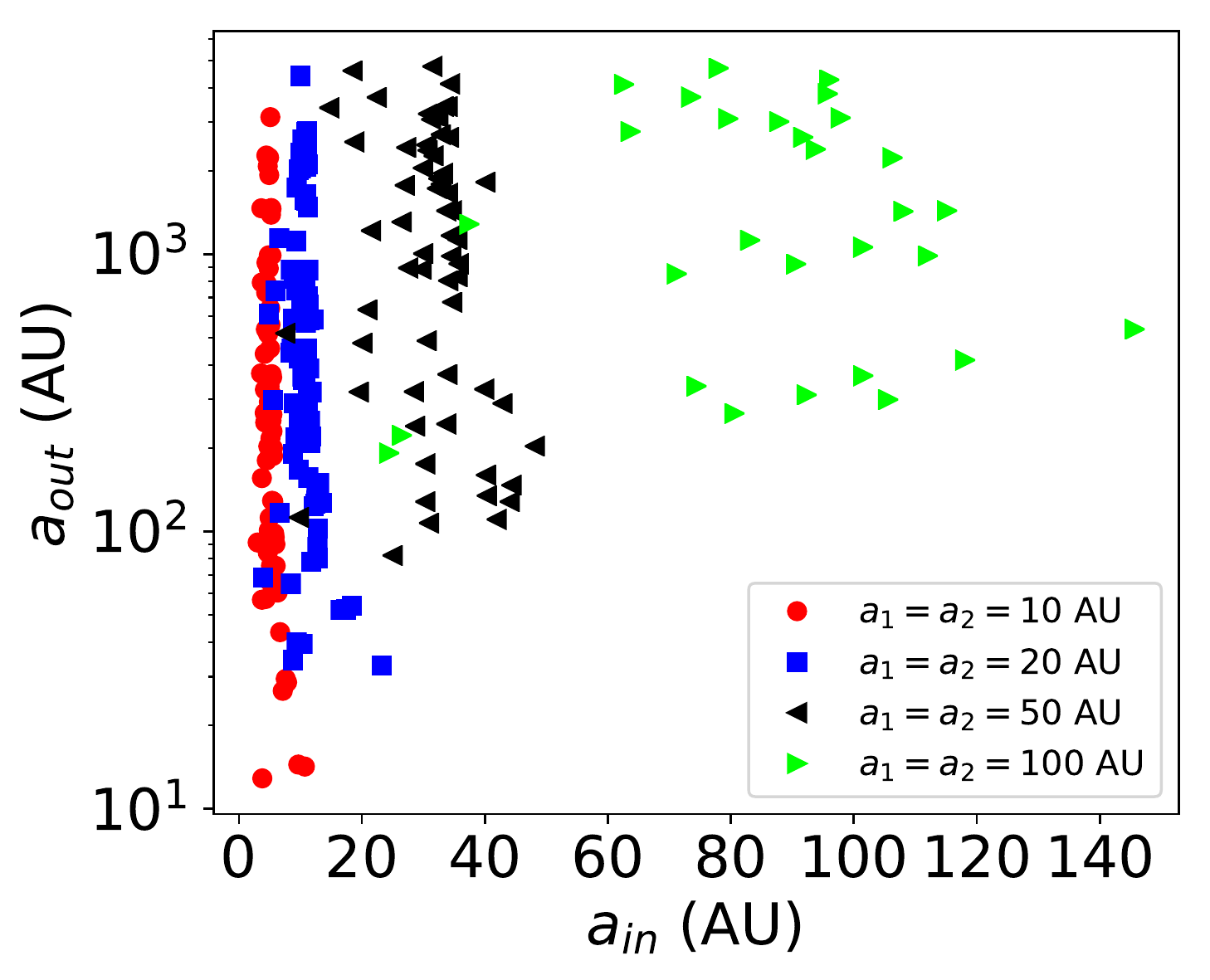}
\includegraphics[scale=0.58]{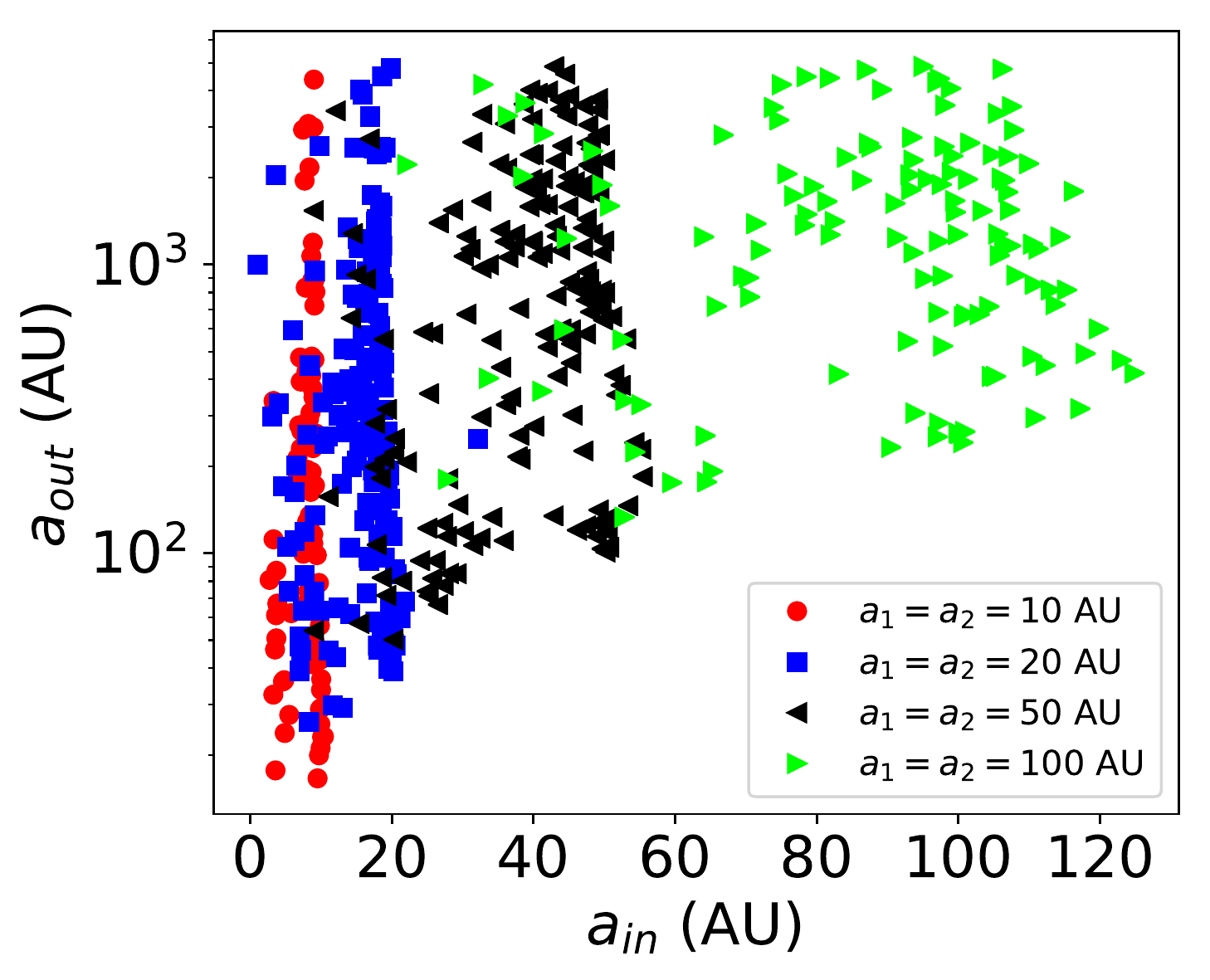}
\includegraphics[scale=0.58]{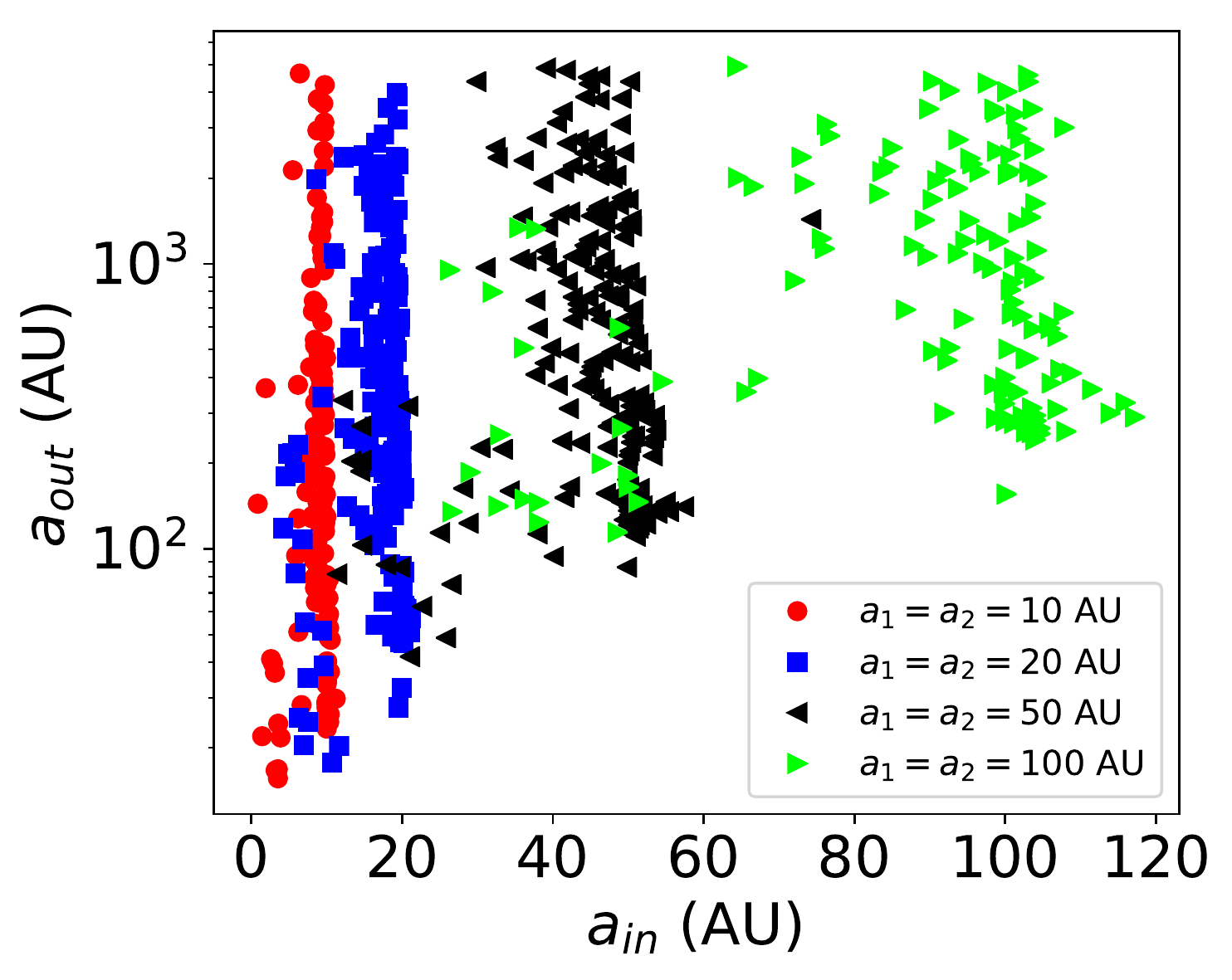}
\caption{Final distribution of inner semi-major axis $a_{in}$ and outer semi-major axis $a_{out}$ of dynamically formed triples with a planet in Model 1 for $m_1=m_2=1 \msun$ (top), $m_1=m_2=3 \msun$ (center) and $m_1=m_2=5 \msun$ (bottom).}
\label{fig:ainaout}
\end{figure}

\begin{figure*} 
\centering
\hspace{0.25cm}$m_1=m_2=1\msun$\hspace{6.25cm} $m_1=m_2=5\msun$
\begin{minipage}{20.5cm}
\includegraphics[scale=0.58]{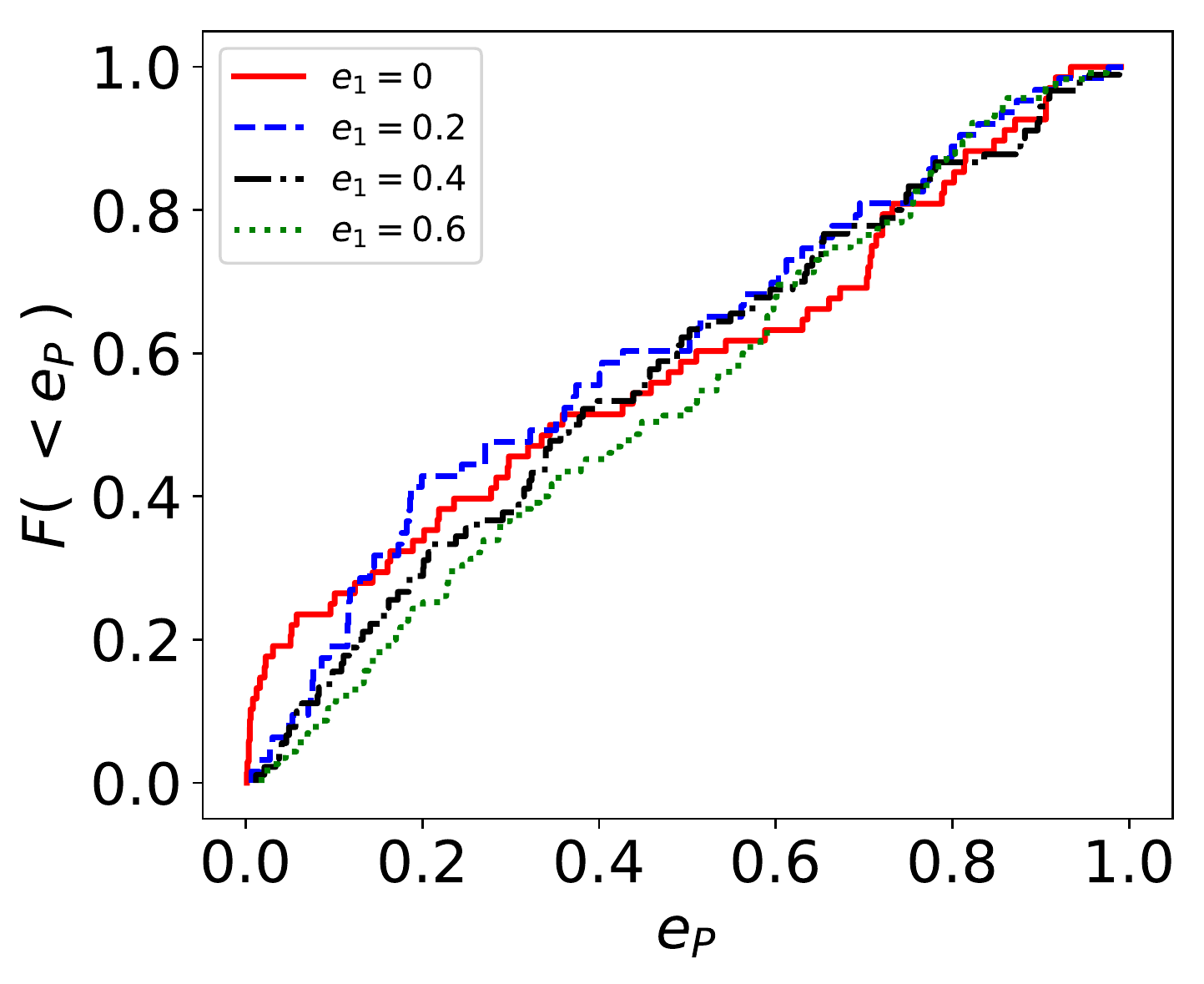}
\includegraphics[scale=0.58]{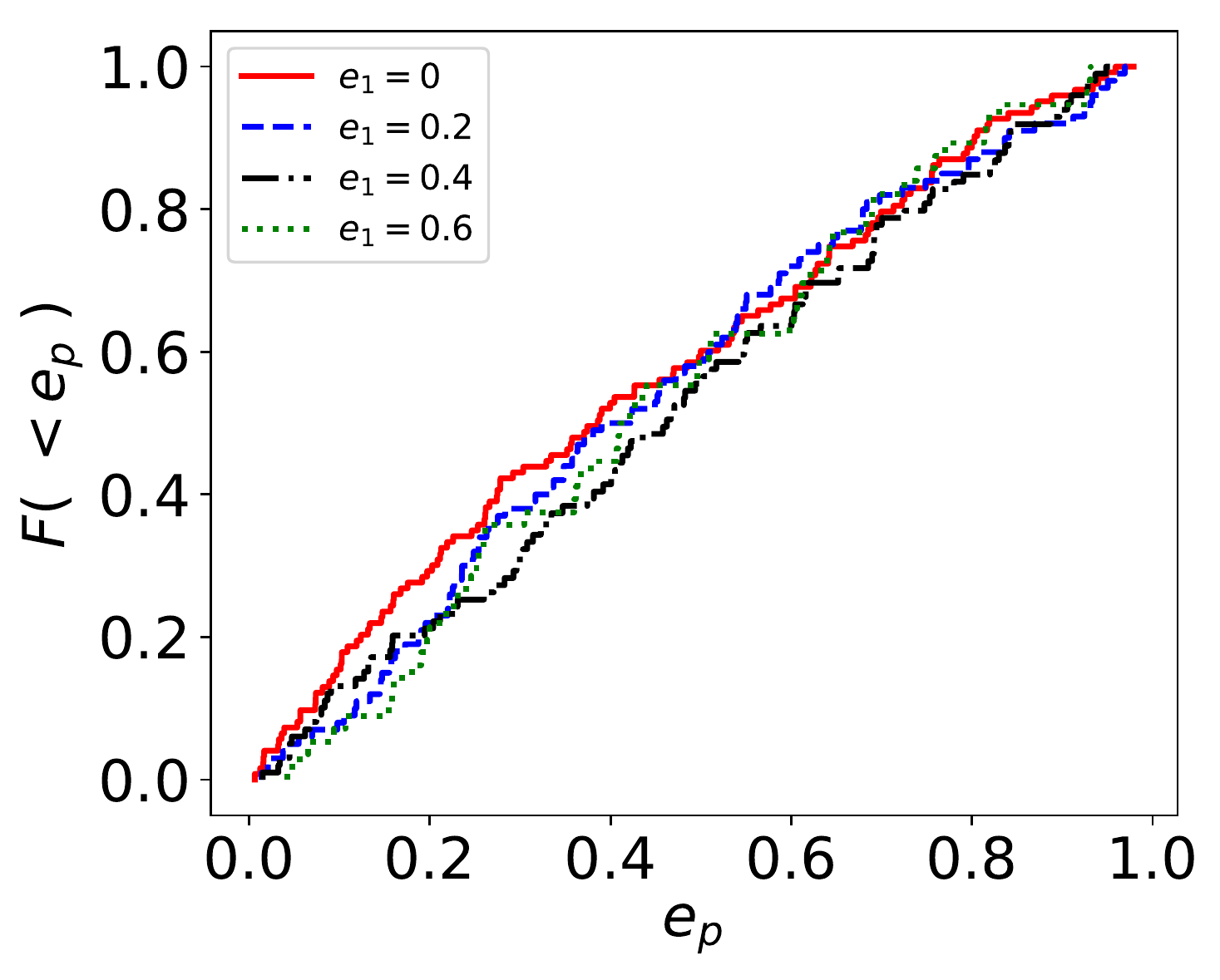}
\end{minipage}
\begin{minipage}{20.5cm}
\includegraphics[scale=0.58]{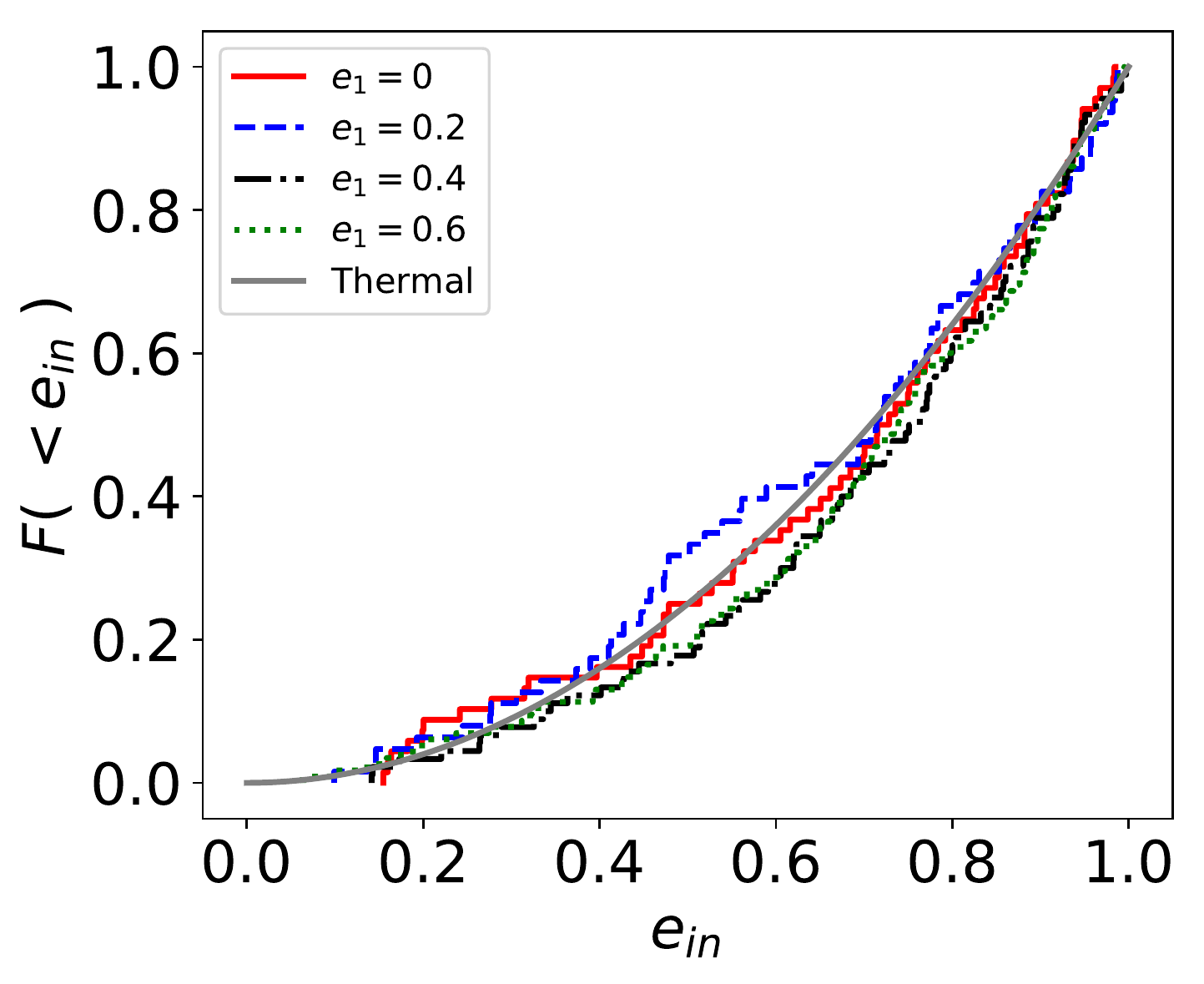}
\includegraphics[scale=0.58]{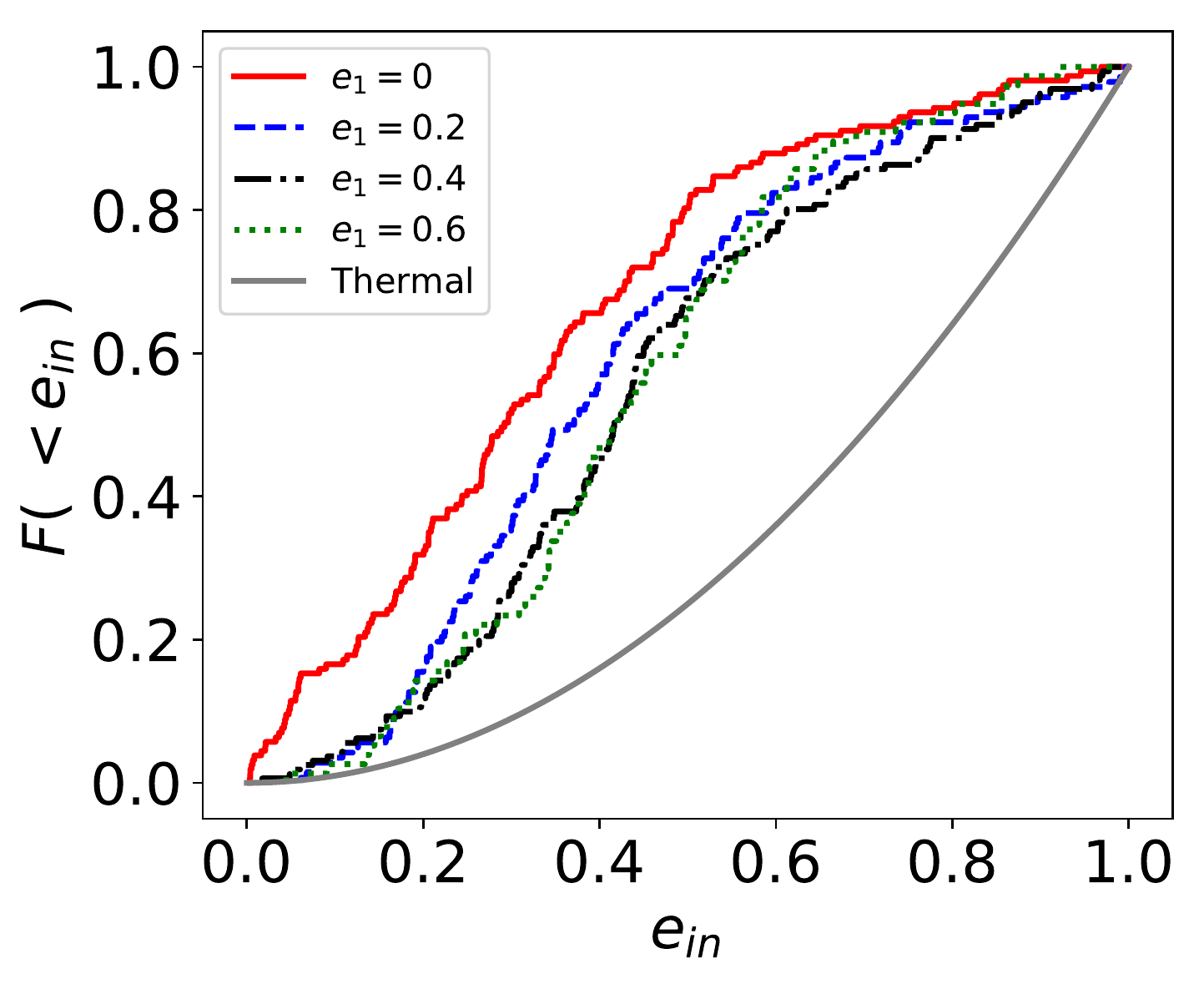}
\end{minipage}
\begin{minipage}{20.5cm}
\includegraphics[scale=0.58]{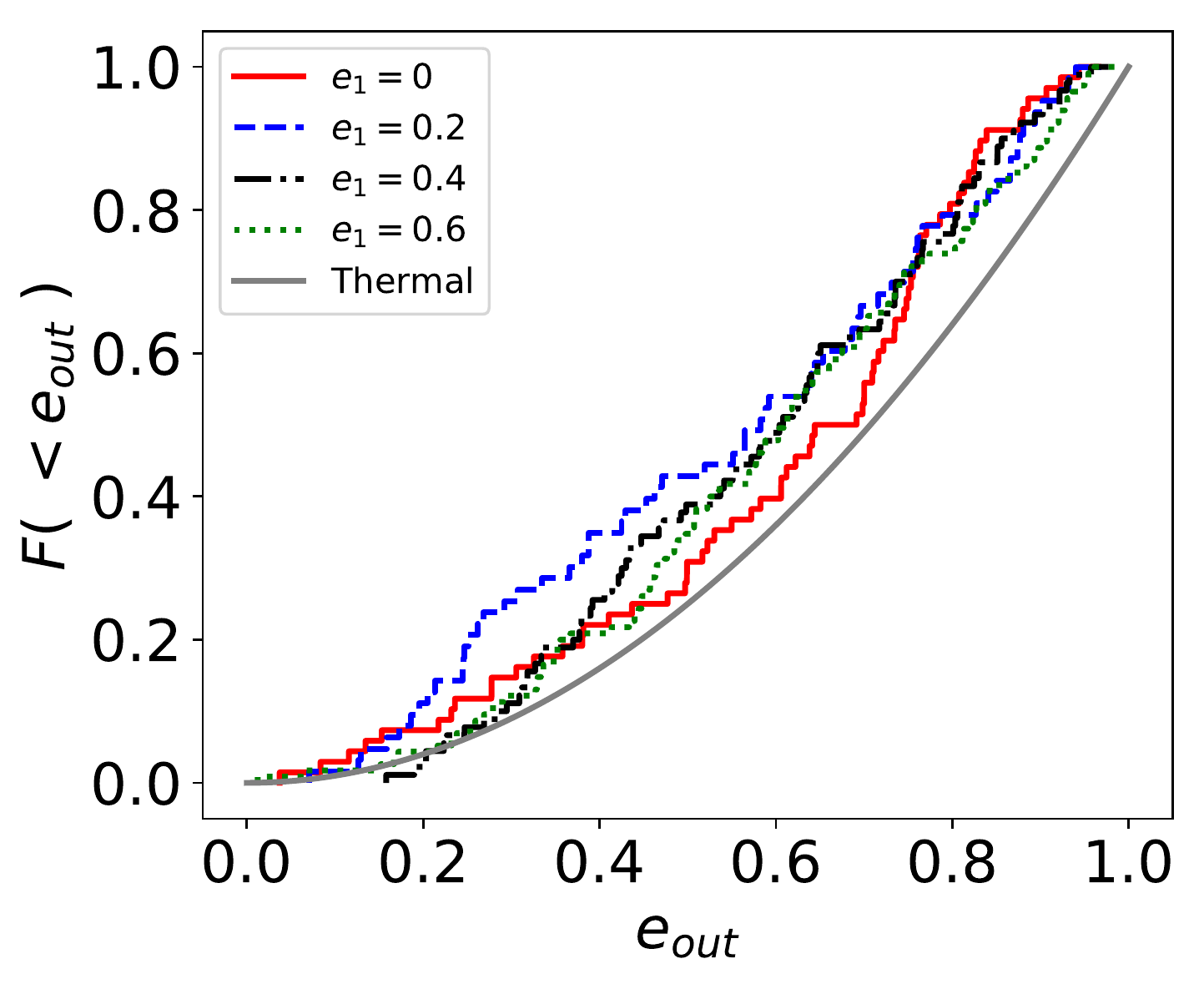}
\includegraphics[scale=0.58]{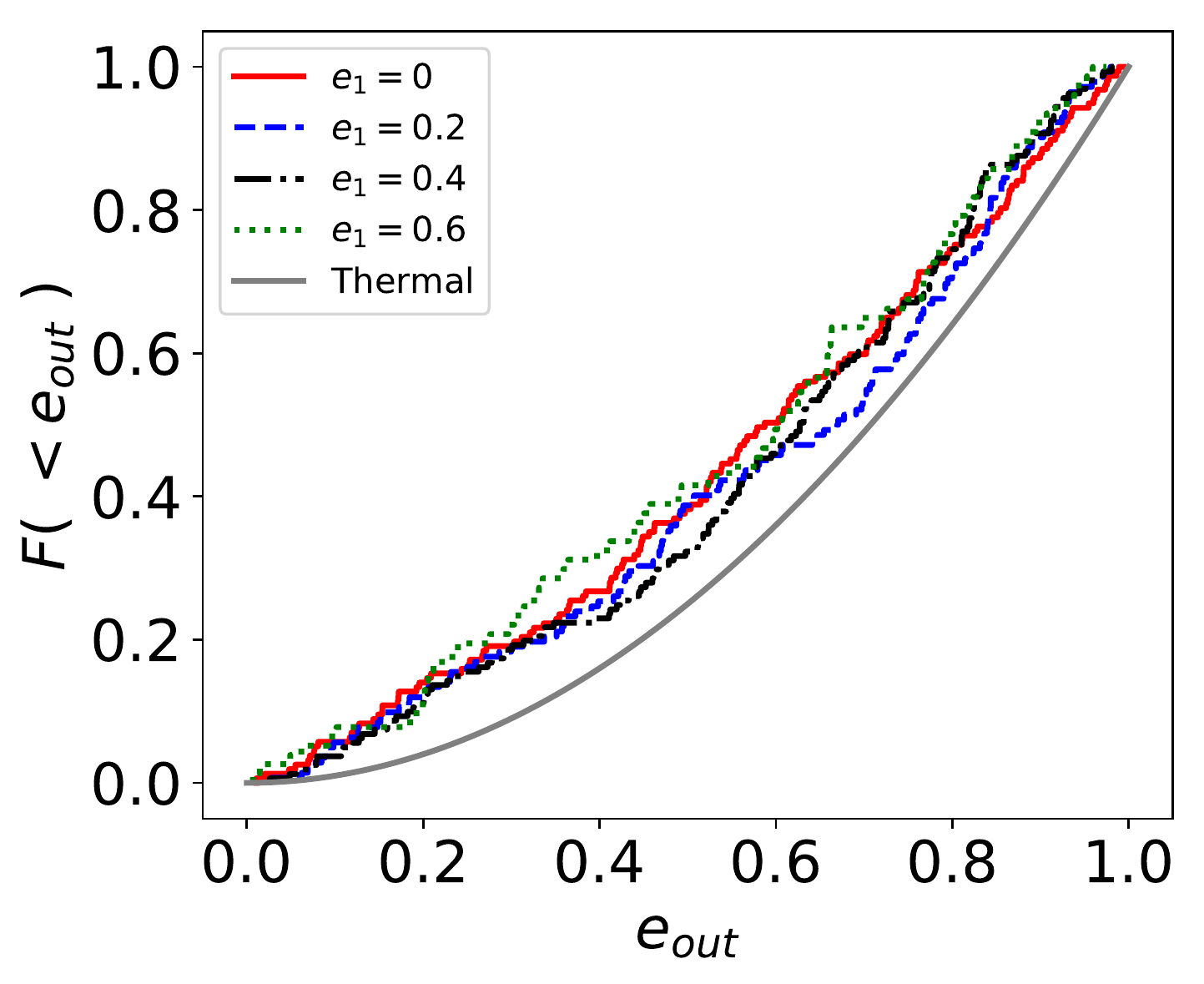}
\end{minipage}
\caption{Final cumulative distribution of the eccentricity of the planet (top), of the eccentricity of the inner orbit $e_{in}$ (centre) and of the eccentricity of the outer orbit $e_{out}$ (bottom) in dynamically formed triple stars that host a planet in Model 2. We also plot the thermal distribution, i.e. $F(e)=e^2$. Left panels: $m_1=m_2=1\msun$; right panels: $m_1=m_2=5\msun$.}
\label{fig:ecc}
\end{figure*}

We consider in total 5 different models, as summarised in Table \ref{tab:models}. In each model, we fix all the parameters, but one, which assumes the different values reported in Table \ref{tab:models}. Additionally, in all the models, we consider different mass ratios $q=M_{12}/M_{34}$ (where $M_{12}$ is the total mass of the initial planet-host binary and $M_{34}$ is the total mass of the second binary) by fixing the mass of the stars in one of the two binaries to $1\msun$ and varying the masses of the stars in the second binary in the range $1-5\msun$. We assign the planet to one of the two stars in the first binary (i.e. either to $m_1$ or $m_2$). In Model 1, we study the effect of the initial binary semi-major axis, by considering $a_1=a_2$ in the range $10$-$100$ AU, while fixing both initial eccentricities $e_1=e_2=0$, the velocity dispersion $\vdisp=3\kms$, and the planet semi-major axis $a_p=1$ AU. To consider how the initial mass ratio $q$ affects the results, we ran Model 1b, where the planet-host binary is less massive than the second binary ($q<1$). In Model 2, we analyse the effect of the initial binary eccentricity ($e_1=0-0.6$ and $e_2=0$) on the fate of the binary-binary scattering, for $a_1=a_2=10$ AU, $\vdisp=3\kms$ and $a_p=1$ AU. In Model 3, we study the role of the initial planetary semi-major axis by varying it in the range $a_p=0.1-1$ AU, while $a_1=a_2=10$ AU, $e_1=e_2=0$ and $\vdisp=3\kms$. In Model 4, we study the effect of the velocity dispersion by considering the range $\vdisp=0.1-3\kms$ for different values of the semi-major axis and $a_P=1$ AU.

Figure \ref{fig:ainaout} shows the final distribution of inner semi-major axis $a_{in}$ and outer semi-major axis $a_{out}$ of dynamically formed triples with a planet in Model 1 for $m_1=m_2=1 \msun$ (top), $m_1=m_2=3 \msun$ (center) and $m_1=m_2=5 \msun$ (bottom). For all mass ratios $q$, different regions of the $a_{in}$-$a_{out}$ plane are populated by triples formed from binaries with different initial semi-major axis. We note that the inner semi-major axis, $a_{in}$, is distributed around a central value that corresponds to the initial binary semi-major axis, while the outer semi-major axis, $a_{out}$, spans a range from a few times the initial binary semi-major axis to a few thousands AU. The plane $a_{in}$-$a_{out}$ offers an important flag for binaries whose interaction leads to a triple star system: the larger the initial binary semi-major axis is, the bigger is the inner semi-major axis of the triple.

In our runs, the planet is always assigned to one of the two stars of the first binary (i.e. $m_1$ or $m_2$), which can be more massive than the second binary, whose masses are fixed at $m_2=m_3=1\msun$. We also ran Model 1b, where the masses of the stars in the first binary were fixed to $1\msun$, with $m_3=m_4$ of either $3\msun$ or $5\msun$ (i.e. mass ratio $q<1$). When two binaries of different masses interact and form a triple, the inner binary in most of the cases is made up of the more massive initial binary with an outer companion captured from the initial less-massive binary. We find that the relative probability of forming a triple with a planet is smaller ($\sim 30-40\%$) in Model 1b than in Model 1. If the planet is initially hosted by the more massive binary (as in Model 1), it will likely orbit one of the stars in the inner binary of the dynamically formed triple. On the other hand, if the planet is initially hosted by the less massive binary (as in Model 1b), it will orbit the outer star of the dynamically formed triple. As noted, the outer star is captured in this case from the initially lighter binary, while the former companion is ejected. Since either the planet-hosting star or the other star in the less-massive binary can be ejected, the probability of hosting a planet in the final triple is smaller in Model 1b than in Model 1.

\begin{figure} 
\centering
\includegraphics[scale=0.58]{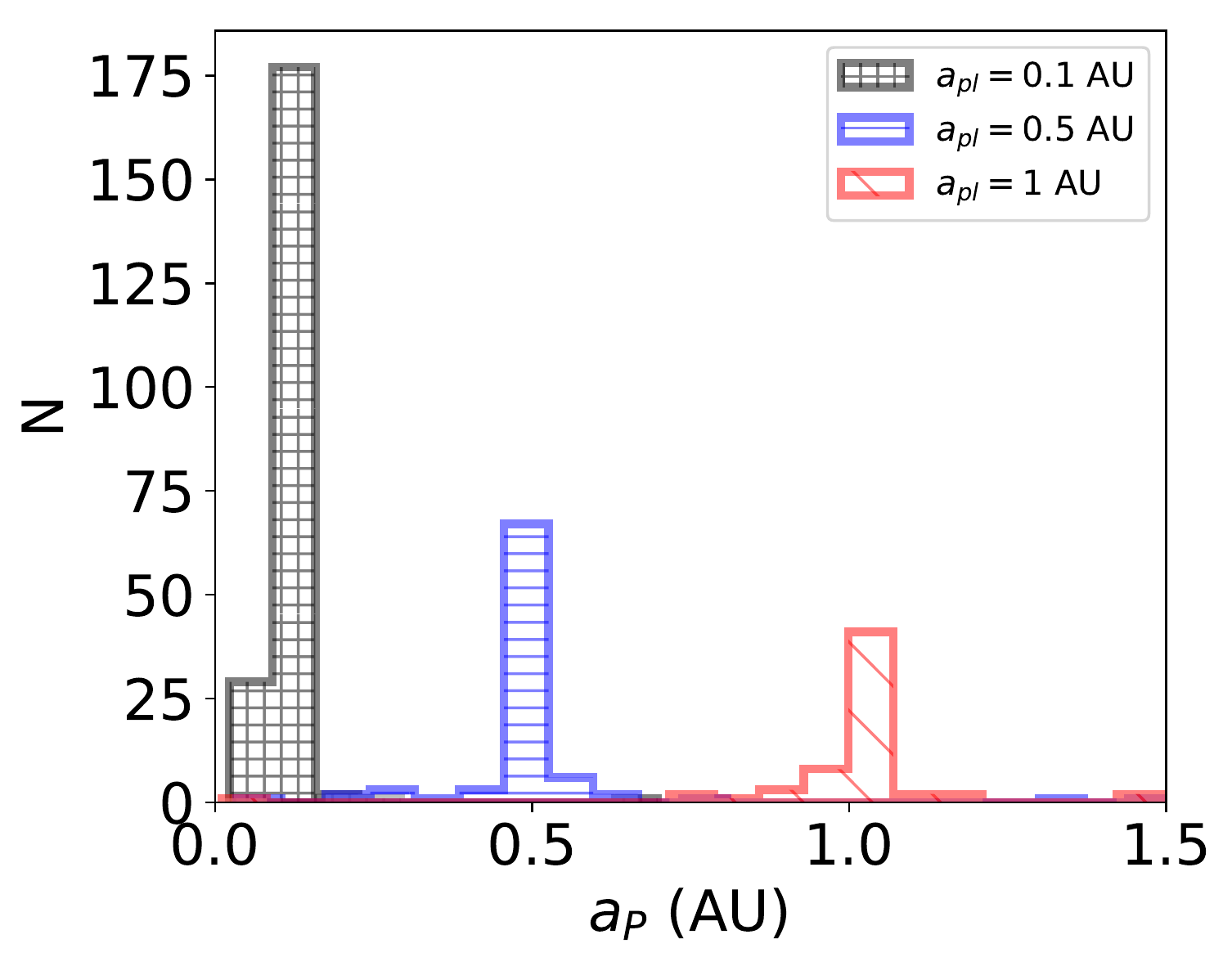}
\includegraphics[scale=0.58]{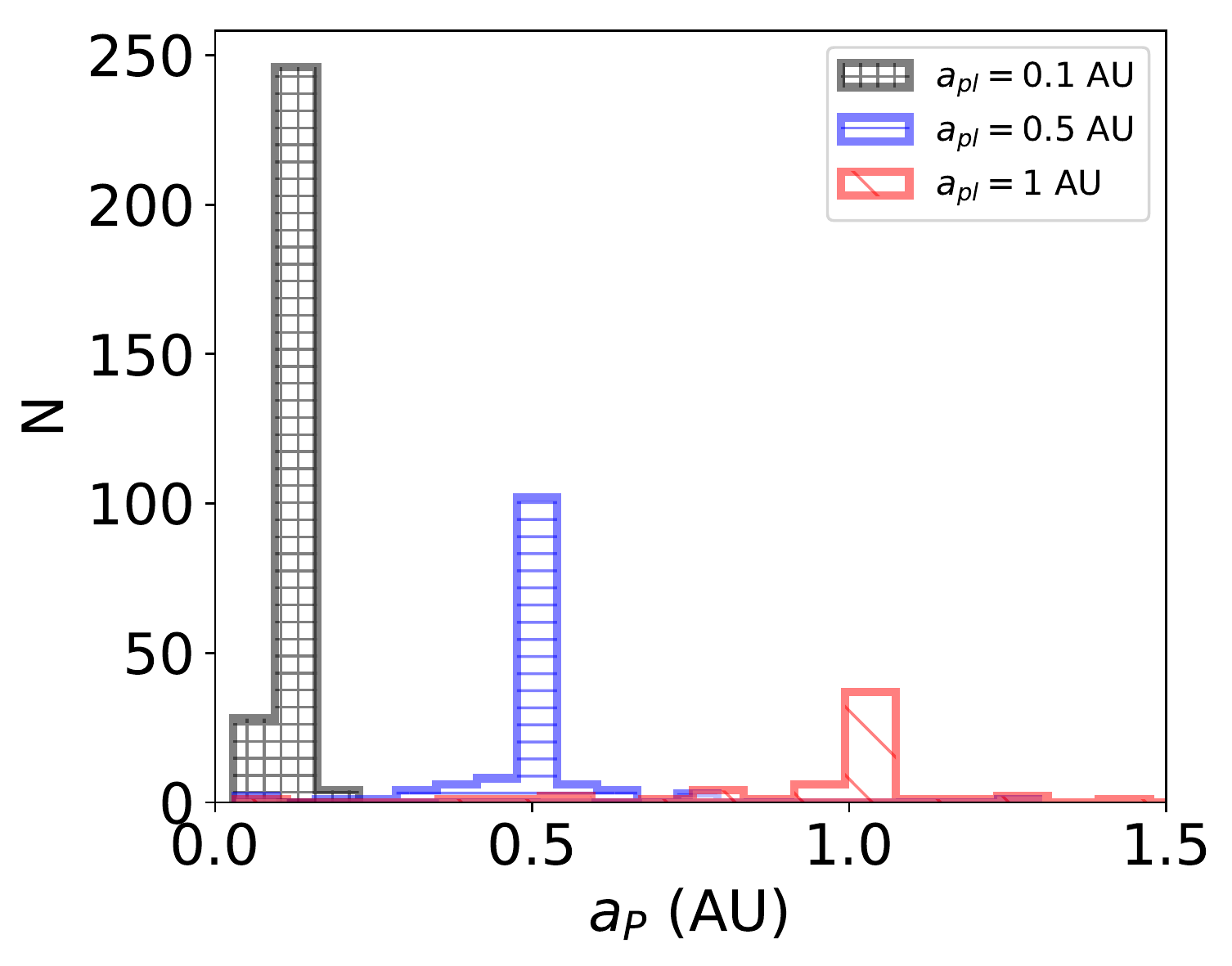}
\includegraphics[scale=0.58]{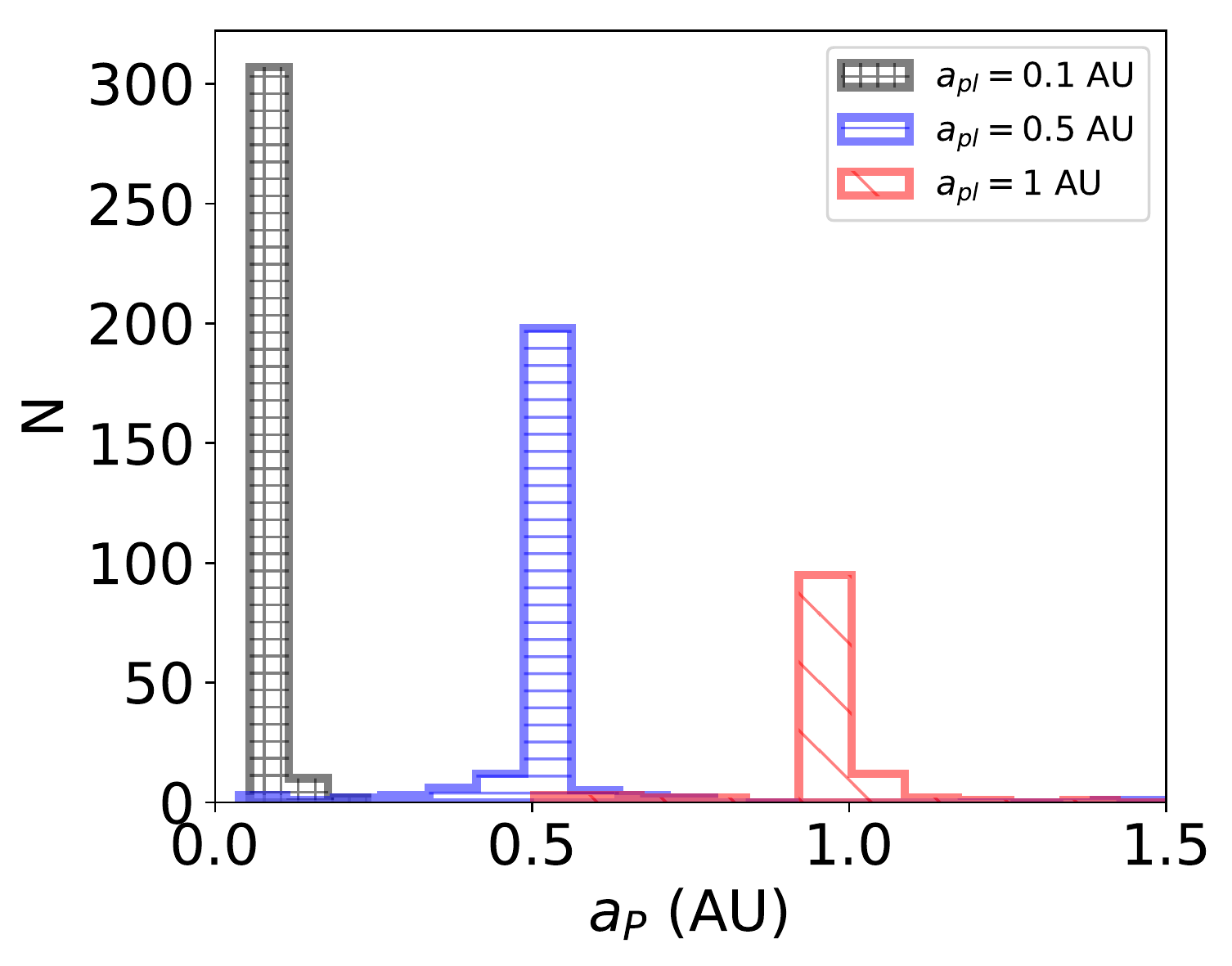}
\caption{Final distributions of planet semi-major axis for $m_1=m_2=1 \msun$ (top), $m_1=m_2=3 \msun$ (center) and $m_1=m_2=5 \msun$ (bottom), in Model 3.}
\label{fig:apl}
\end{figure}

\begin{figure} 
\centering
\includegraphics[scale=0.58]{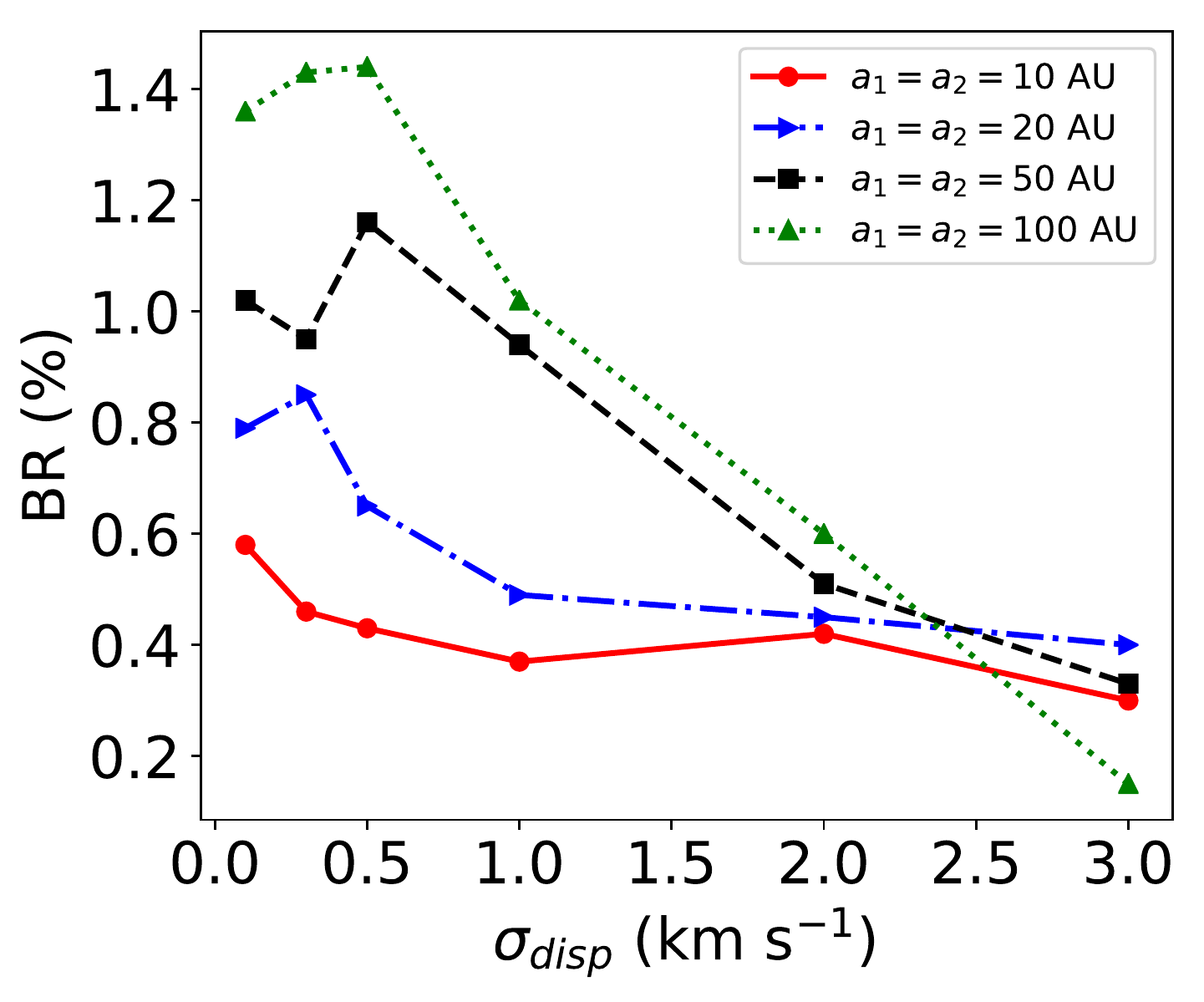}
\caption{Branching ratios (probability of outcome in percent) of triples with a planet as function of the velocity dispersion and different semi-major axis in Model 4.}
\label{fig:br_vdisp}
\end{figure}

We show the cumulative distributions of the eccentricity of the planet, of the inner orbit and outer orbit of the dynamically formed triples hosting a planet in Fig. \ref{fig:ecc}, both for $m_1=m_2=1\msun$ (left panels) and for $m_1=m_2=5\msun$ (right panels). Independently on the mass ratio of the involved binaries and initial binary eccentricity, the cumulative distribution of planet eccentricities in the final triples is roughly linear. We note that we removed from the analysis planets that end with an eccentric orbit whose pericentre is smaller than the sum of the star and planet radii. While for $q=1$ this fraction is quite small ($\sim 5-7$\%), for $q=5$ it is roughly a quarter of all the systems. The initial eccentricity also plays an important role: the larger $e_1$, the larger is the fraction of planets that collide with their host star. The effect of the mass ratio is clear for the distribution of the eccentricity of the inner orbit, which is thermal (i.e. $F(e)=e^2$) in the case $q=1$, and deviates significantly from it for the case $q=5$. Finally, the eccentricity of the outer orbit, $e_{out}$, is always thermally distributed independently of the mass ratio. 

In Model 3, we study the role of the initial planetary semi-major axis, $a_P$. Figure \ref{fig:apl} reports the final distributions of the semi-major axis of planets in dynamically formed triples for $m_1=m_2=1 \msun$ (top), $m_1=m_2=3 \msun$ (center) and $m_1=m_2=5 \msun$ (bottom), as function of the initial $a_P$. The final semi-major axis is mostly determined by the initial semi-major axis, having a distribution peaked at the initial $a_P$ with a spread that is larger for a larger initial semi-major axis. Changes of the planet semi-major axis from its initial values may be due to both the intrinsic evolution of the binary and the binary-binary encounter. The smaller the initial $a_P$ is, the larger is the number of planets that merge with their parent star (resulting in removal from the plotted distributions), but also the larger is the probability of forming triples with a planet. Figure \ref{fig:apl} shows also the number of planets that are in dynamically formed triples out of the same number of total simulations (25k for each combination of $q$ and $a_P$). We find that the number of systems is larger when the mass ratio is larger. The more likely output of a massive binary interacting with a lighter binary is either the ionization of the second binary or the capture of one of the stars in the lighter binary, that forms a bound triple with the more massive binary. The larger the mass ratio, the larger is the probability to form a bound triple by capturing a third companion. 

Figure \ref{fig:br_vdisp} reports the branching ratios (BRs), i.e. the probability of the outcome, for dynamically formed triples as function of the velocity dispersion of the binaries, for different initial semi-major axis ($a_1=a_2$; Model 4). The velocity dispersion depends on the environment where binary-binary events take place, namely the host star cluster. Open cluster environments have typical velocity dispersions $\vdisp\approx 1\kms$, while globular clusters have larger velocity dispersions, $\vdisp\approx 10\kms$. We find that the maximum likelihood of forming triple systems hosting a planet is achieved when $\vdisp\lesssim 1\kms$, i.e. the typical velocity dispersion of open clusters. In this regime, we find that the probability that a dynamically formed triple is a planet-hosting triple is of the order of $\sim 0.5-1.5\%$. As discussed previously in Fig. \ref{fig:apl}, larger mass ratios increase this probability and could enhance the fraction of dynamically formed planet-hosting triples up to $\sim 3\%$. 

\section{Conclusions}

About two dozens of triple star systems have been observed to host a planet. Several follow-up observations on the stellar multiplicity of planet-host stars have revealed that $\sim 2.5\%$ of planets are in triple and multiple systems \citep{rag06,mug09,roe12}. In this paper, we have considered a dynamical origin for planets in triple systems as a consequence of binary-binary scatterings in star clusters \citep{lei13}, as firstly proposed by \citet{port05}. For the first time, we have included directly the planet in the scattering experiments.

We have examined different masses, mass ratios, orbital semi-major axis and eccentricities for the stars in the binaries that undergo binary-binary interactions. We have shown that different regions of the $a_{in}$-$a_{out}$ plane are populated by dynamically formed triples for all the mass ratios, with the inner semi-major axis distributed around a central value that corresponds to the initial binary semi-major axis, while the outer semi-major axis spans the range from a $\sim$ few times the initial binary semi-major axis to a $\sim$ few thousands AU. Moreover, while the final eccentricity of the outer orbit in the formed triples is thermally distributed, the inner orbit eccentricity deviates from it when the initial binaries have different masses.

We have also analysed the role of the initial planet semi-major axis. We found that the smaller the initial $a_P$ is, the larger is the number of planets that merge with their parent star, but also the larger is the probability of forming planet-hosting triples. Moreover, large mass ratios imply a larger number of hierarchical triples.

Finally, we studied how the local environments affect the dynamical formation of planets in triple systems by considering different velocity dispersions $\vdisp$. We found that the maximum likelihood of forming triple systems is for $\vdisp\lesssim 1\kms$, i.e. the typical velocity dispersion of open clusters. For this case, we calculated a formation probability of the order $\sim 0.5-3\%$.

Our results suggest that binary-binary encounters are a viable way of creating planet-hosting triple systems, in the mass range studied in this paper. We note that in our models we always assume that only one of the stars in the two binaries hosts a planet. However, more than one star could host a planet, or even a multi-planetary systems, thus enhancing the rates. At the same time, some binaries might not host planets at all, thus reducing the formation probability. Also, the long-term evolution of these systems is crucial in understanding how many of them are stable and observable. Finally, we also note that our results are connected to the origin and stability of planets in S-type orbits in binary stars, whose exact relation to the model presented in this paper deserves future work \citep{thh15}.

The origin of planets in triple systems is still highly debated. The question which of the proposed formation mechanisms (binary-binary encounters, primordial triple formation, capture of planets by a triple, and capture of stars with planets by binaries) is the dominant one is still far from having an answer. The recently launched TESS is expected to discover a larger sample of planets in triple systems than currently known. Future data by the James Webb Telescope, along with upcoming exoplanets missions like PLATO and CHEOPS, may shed additional light on the triple star systems in star clusters and their exoplanet population.

\section*{Acknowledgements}
This research was supported in part by an ISF and an iCore grant, as well as a grant from the Breakthrough Foundation. GF is supported by the Foreign Postdoctoral Fellowship Program of the Israel Academy of Sciences and Humanities. GF also acknowledges support from an Arskin postdoctoral fellowship and Lady Davis Fellowship Trust at the Hebrew University of Jerusalem. GF acknowledges hospitality by the Institute for Theory and Computation at the Harvard-Smithsonian Center for Astrophysics, where the early plan for this work was conceived. Simulations were performed on the \textit{Astric} cluster at the Hebrew University of Jerusalem.

\bibliographystyle{mn2e}
\bibliography{biblio}
\end{document}